# Low levels of sulphur dioxide contamination of Venusian phosphine spectra

Jane S. Greaves,[1]★ Paul B. Rimmer,[2,3,4] Anita M. S. Richards[5], Janusz J. Petkowski[6], William Bains[1,6], Sukrit Ranjan[7], Sara Seager[6,8,9], David L. Clements[10], Clara Sousa Silva[11] and Helen J. Fraser[12]

[1]School of Physics & Astronomy, Cardiff University, 4 The Parade, Cardiff CF24 3AA, UK
[2]Department of Earth Sciences, University of Cambridge, Downing Street, Cambridge CB2 3EQ, UK
[3]Cavendish Astrophysics, University of Cambridge, JJ Thomson Avenue, Cambridge CB3 0HE, UK
[4]MRC Laboratory of Molecular Biology, Francis Crick Ave, Trumpington, Cambridge CB2 0QH, UK
[5]Jodrell Bank Centre for Astrophysics, Department of Physics and Astronomy, The University of Manchester, Manchester, UK
[6]Department of Earth, Atmospheric, and Planetary Sciences, Massachusetts Institute of Technology, 77 Mass. Ave., Cambridge, MA, 02139, USA
[7]Northwestern University, 1800 Sherman Avenue 8041, Evanston, IL 60201, USA
[8]Department of Physics and Kavli Institute for Astrophysics and Space Research, Massachusetts Institute of Technology, 77 Mass. Ave., Cambridge, MA, 02139, USA
[9]Dept. of Aeronautics and Astronautics, Massachusetts Institute of Technology, 77 Mass. Ave., Cambridge, MA, 02139, USA
[10]Department of Physics, Imperial College London, South Kensington Campus, London SW7 2AZ, UK
[11]Harvard-Smithsonian Center for Astrophysics, Observatory Building E, 60 Garden St, Cambridge, MA 02138, USA
[12]School of Physical Sciences, The Open University, Walton Hall, Milton Keynes MK7 6AA, UK

**ABSTRACT**

New analysis is presented of the 1.1 mm wavelength absorption lines in Venus' atmosphere that suggested the presence of phosphine. We retrieve a sulphur dioxide observation from the JCMT archive that was simultaneous within a few days of the $PH_3$ 1-0 spectrum obtained in June 2017, and demonstrate via a radiative transfer calculation that contamination of $PH_3$ by $SO_2$ was ≈10%. We also present ALMA 2019 spectra of $PH_3$ 1-0 and an $SO_2$ transition acquired simultaneously, and infer that $SO_2$ line-contamination was ≲2% (for the least-noisy half of the planetary disc). The contamination-subtracted ALMA and JCMT spectra (of 6-8 sigma confidence) are now consistent with similar absorption-depths at the two epochs. The two values span -1.9(±0.2) $10^{-4}$ of the continuum signal (which was re-estimated for ALMA), albeit for differing planetary areas. This suggests that the abundance attributed to phosphine in Venus' atmosphere was broadly similar in 2017 and 2019.

## 1 INTRODUCTION

There is presently little consensus about the possible presence of phosphine, $PH_3$, in Venus' atmosphere. This trace-gas would be very unexpected for an oxidised planet, but could come from unknown chemical routes (Bains et al. 2021a), or even be a biological by-product. Ultra-dry/acidic cloud conditions would be uninhabitable by any known life (Seager et al. 2021a, Hallsworth et al. 2021), but neutralisation of some of the acid in droplets could provide better habitats (Bains et al. 2021b). In Greaves et al. (2020) we presented 1.1 mm-wavelength absorption features interpreted as from $PH_3$ in Venus' atmosphere, owing to close alignment with the 1-0 rotational-transition frequency. Mogul et al. (2021) then retrospectively attributed to phosphine a signal from the Pioneer-Venus mass spectrometer LNMS, which sampled the clouds in 1978. All three of these signals were at ≳5σ confidence after careful re-processing. Further searches for phosphine are currently difficult — shorter-wavelength absorption only traces above-cloud layers (Trompet et al. 2021), and further in-situ sampling awaits new missions. Planned future instruments include the mass spectrometer VMS on NASA's

DAVINCI, and a balloon-borne tunable laser spectrometer as part of the Venus Life Finder programme (Seager et al. 2021b).

Critiques of our results have posed three main questions: (1) are the 1.1 mm features robust? (2) could they be due to $SO_2$ absorption rather than $PH_3$? (3) do the line-widths observed suggest the absorbing gas is at high altitude, where $PH_3$-presence is problematic as it should be rapidly photo-chemically destroyed? Regarding (1), Greaves et al. (2021a,b) presented new data-processing addressing technical issues (Villanueva et al. 2021) and statistics on false-positives (Snellen et al. 2020; Thompson 2020). We here expand on some of the technical issues in (1), and then focus on (2), with new analysis quantifying the level of $SO_2$ contamination of the 1.1 mm features we attributed to $PH_3$. We find that $SO_2$ contamination levels are insignificant compared to those from ad-hoc models (Akins et al. 2021; Lincowski et al. 2021). Addressing (3) will require new observations and better laboratory data, as inversion of line-shapes to altitude-profiles of abundance needs lower-noise spectra and a $PH_3$-$CO_2$ broadening-coefficient that improves on current theoretical estimates.

The 1.1 mm data discussed here include: from the James Clerk Maxwell Telescope (JCMT), a centred-weighted whole-planet spectrum observed over a week in June 2017, and from the Atacama Large Millimeter/submillimeter Array (ALMA), a datacube from one morning in March 2019, where the planet was spatially resolved. With ALMA, issues with spatial filtering (Villanueva et al. 2021, Akins et al. 2021, Greaves et al. 2021a) resulted in least smooth regions, e.g. the limb, being the best-sampled. Full details are given in Table 1.

★ E-mail: greavesj1 at cardiff.ac.uk

**Table 1.** Details of the observations made with JCMT and ALMA. The JCMT data are single-dish spectra covering the planetary areas listed. The ALMA interferometer was configured with baselines between antennas of 15-314m, but data shown here exclude baselines of <33m to reduce noise; the largest angular scale detected is then smaller than Venus.

| facility | project id | dates | integration times | frequency centres and bands | planetary area covered |
|---|---|---|---|---|---|
| JCMT | S16BP007 | 2017 June 16/14/12/11/9 | 30 minutes ×10 | 266.946 GHz, 250 MHz band | 75% (at ≥ half-power of beam) |
| JCMT | M17AP081 | 2017 June 6, May 26/17 | 25-40 minutes /day | 346.590 GHz, 1 GHz band | ~30% (at ≥ half-power of beam); also offset beams viewing limb (see text) |
| ALMA | 2019.A.00023.S | 2019 March 5 | 3.5 hour block | 266.1 & 266.9 GHz, with 1875 & 117 MHz bands | 100% at ≥70% power, but largest scale detected ≈ Venus-radius (see text) |

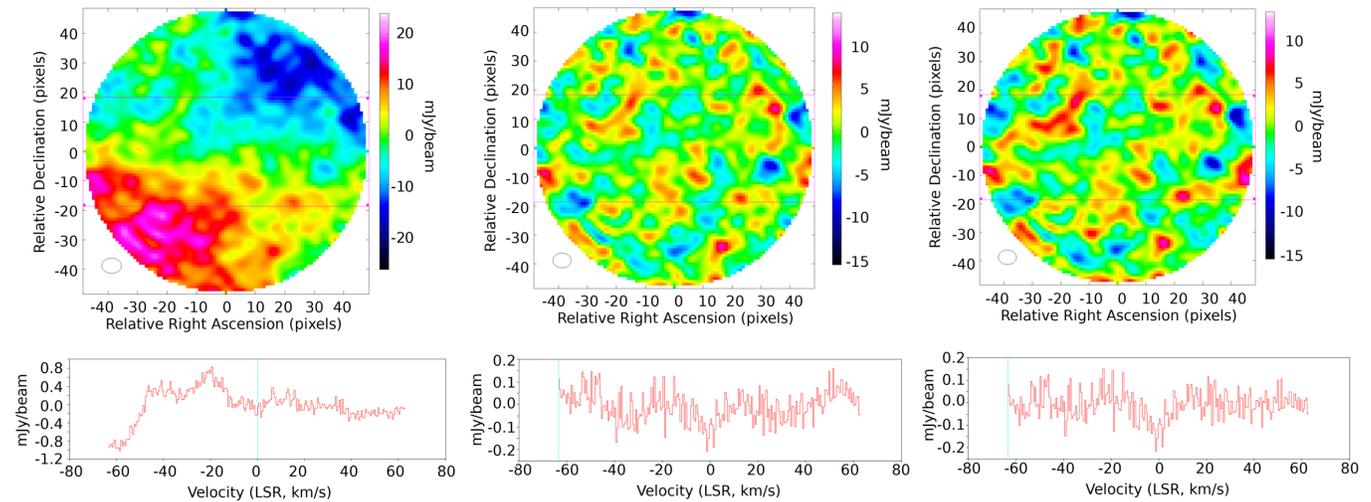

**Figure 1.** Final stages of the ALMA processing, as performed for this project. Each top panel shows the zero-velocity slice of the calibrated PH$_3$ 1-0 datacube. From left to right: Venus continuum-subtracted cube using all antenna baselines (15-314m) and subtracting only a 1st-order image-plane fit; using only baselines of 33-314m and with a 1st-order image-plane fit; using only baselines of 33-314m and with a 6th-order image-plane fit. The axes are RA,Dec with 0.16 arcsec pixels. Flux scales are Jy per beam with beam size shown by the ellipses at lower-left of each panel. The lower panels show the corresponding spectrum from each of the identical boxes (of 97×37 pixels) outlined in purple in the upper panels. The axes are flux in mJy/beam against radio-definition velocity with respect to Venus. Spectral ripple is more suppressed at velocities away from zero with the higher-order image-plane fit, and so the lower-right spectrum has the lowest residuals. The box in which extraction was performed was chosen to illustrate a good balance of planetary coverage and ripple cancellation.

## 2   DATA PROCESSING

### 2.1   ALMA

The original ALMA processing suffered from several issues in bandpass calibration and flux-scaling. The observations were non-optimal in calibration, being exploratory in high dynamic range for line:continuum. (For comparison, only one similar ALMA dataset on Venus has been published – by Encrenaz et al. (2015); Vandaele et al. (2017); Piccialli et al. (2017) – with only around half the observing time and one-third of the antennas employed for our project, or $\sim \sqrt{6}$ times the sensitivity.) After the observatory (JAO/ESO) re-calibrated and re-released our Venus data in late 2020, the line-shapes were seen to be improved – i.e. were more "V-shaped" as expected for a narrow line with faint pressure-broadened wings, after the removal of the more "bowl-shaped" artefacts. However, there are still caveats on recoverable planetary area and line:continuum fidelity, as discussed below.

Going beyond the JAO-released re-processing, Greaves et al. (2021a) excluded short antenna-baselines, and applied low-order image-plane fits. The first procedure removes at an early stage the data that contribute the most spectral "ripple" and noise – Greaves et al. (2021a) demonstrated that including the shorter antenna-baselines increases the noise by a factor ≈2.5. However, without the short baselines, the largest angular scale that is reliably imaged is 4.3 arcsec and the largest detected is 7.1 arcsec (Greaves et al. 2020). Hence, as Venus then subtended 15.2 arcsec, most of the centre of the planet is heavily downweighted, with most of the line signal coming from an annulus inwards of the limb. The second additional procedure involved subtracting image-plane fits, which were of 6$^{th}$-order for the narrowband data including PH$_3$ 1-0, and of 3$^{rd}$-order for the simultaneous wideband data covering a strong SO$_2$ transition. This procedure removes the need for any trendlines (which could be user-biased) to be fitted to spectra extracted from the final datacubes. The orders were chosen as the lowest that could effectively flatten the spectral baselines. Figure 1 illustrates the improvements to spectral quality from these two procedures. Spectra can then be extracted over broad planetary areas, where residual spectral ripples tend to self-cancel[1], as shown by the lower-right panel of Figure 1. Line:continuum ratios are finally obtained via a similarly-processed continuum signal extracted from the same area. This overcomes the difficulties in appropriately calibrating models of Venus, but involves the assumption that line and continuum sources are similarly spatially distributed, as discussed below. The spectra presented here use the processing scripts first published in Greaves et al. (2021a), but here we have recovered phosphine over a larger planetary area.

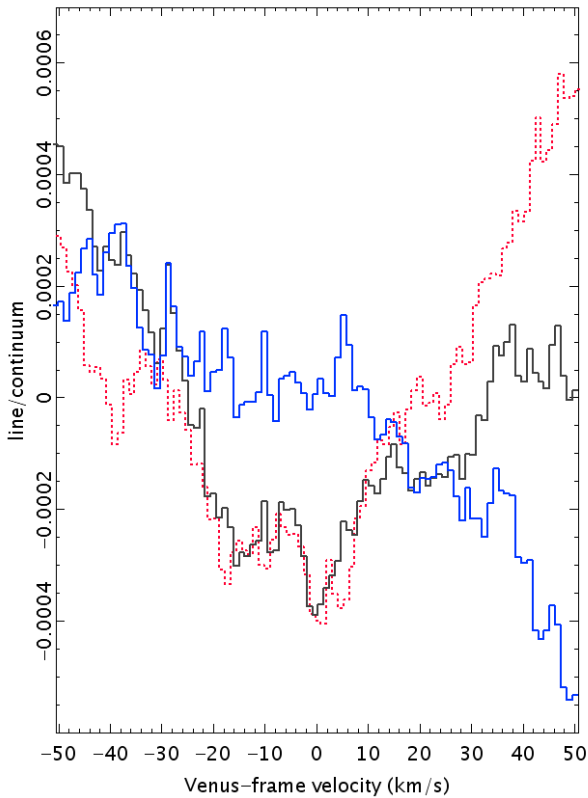

**Figure 2.** Additional processing of the ALMA wideband SO$_2$ spectrum. The black histogram is a spectral cut around the SO$_2$ 13(3,11)-13(2,12) frequency, for the half-planet area of Section 4.1. The dotted red histogram is a "template" line-free section of the same spectrum, after an applied shift of +145 km/s. Subtracting the template from the line-region produces the thick blue histogram, representing the SO$_2$ spectrum with reduced ripple.

The ALMA wideband data were useful as they serendipitously covered several transitions of SO$_2$, with none of these frequencies covered in the narrow-band mode. One post-processing step is also applied here, as the wideband data suffered particularly from spectral ripple (not improved by higher-order image-plane fitting). These wideband ripples are notably quasi-periodic on the frequency axis, and so we experimented with self-cancellation around the strong SO$_2$ transition at 267.537 GHz. A "template" line-free section of spectral baseline was identified around -145 km/s, which has similar structure to the line region, and so can be shifted in velocity and subtracted – see Figure 2. This procedure partially cancelled the largest ripples around the SO$_2$ line.

## 2.2 JCMT

The JCMT PH$_3$ spectra of Greaves et al. (2020) were acquired with bandwidth insufficient to cover any strong SO$_2$ transitions. However, we have now found nearly contemporaneous monitoring of SO$_2$ in the JCMT public archive. The JCMT's HARP camera-spectrometer monitored Venus from March to June 2017 (most relevant dates listed in Table 1), with the 6 June 2017 data acquired very shortly before our PH$_3$ observations on 9-16 June 2017. These raw archival spectra from HARP (0.98 MHz spectral resolution) were downloaded, and one-(x,y)-pixel datacubes were created with the 'makecube' task in SMURF[2], with output spectra viewed and manually corrected for Venus-Earth velocity-differences (>10 km/s) within SPLAT[3]. The resulting SO$_2$ 19(1,19)-18(0,18) (346.65217 GHz) lines have high S/N but some spectral ripple is evident.

HARP observed at 346.6 GHz with a telescope beam of ≈14 arcsec, while Venus' diameter on 6 June 2017 was 23 arcsec. The planet was thus undersampled by the on-source receptor (Table 1), but HARP receptors at 30 arcsec-offsets saw the planetary limb in the first sidelobe. We tested co-adding these offset line:continuum spectra with the planet-centre spectrum, with various offset-weights up to 50%, and found only up to ~10% changes in net SO$_2$ line-depth. The planet-centre spectrum (with the best S/N) is thus adopted as representing net planetary SO$_2$.

## 3 MODELLING

Here we model SO$_2$ spectra using the simplified radiative-transfer code of Greaves et al. (2021a) — the implementation is described here in Appendix A, where a link is provided to the code. The model uses the temperature-pressure profile of the Venus International Reference Atmosphere (VIRA), with details of the altitude and diffusion profiles listed in Greaves et al. (2020). Spectra are produced in absorption against a "quasi-"continuum formed by broad blended emission features of molecules deeper in the atmosphere. Appendix A below describes how the SO$_2$ line profiles are calculated. The aim here is for the results from simple pencil-beam calculations to be reproducible and to use transparent methods. We have neglected convolution with the telescope beam, as these effects should be similar for line and continuum. (There could be some differences if an absorbing molecule has a spatial distribution unlike the molecules producing the quasi-continuum – for example, if absorption favours the day- or night-side – however, prior ALMA data do not indicate any such clear trends (e.g. see Figures 15-16 in Piccialli et al. 2017).) All the lines we model have very low optical depths (line/continuum « 1), and so trace all altitudes above where the quasi-continuum is opaque, but there is a strong weighting function according to local temperature gradient (e.g. Figure 7 of Piccialli et al. 2017). We also neglect the decrease in absorption towards the planetary limb, having described above the JCMT case where this is found to influence the whole-planet spectrum by only ~10%.

The abundances inferred here are intended to be *representative*

---

[1] The combination of signals from pairs of telescopes in radio interferometry produces complex visibilities which can be expressed as a combination of sinusoidal functions, generally producing in the image plane amplitude errors that are symmetric and phase errors that are anti-symmetric about the image centre (Taylor et al. 1999), as is the case here. Hence, the errors tend to cancel for a selected region that is symmetric about the centre.

[2] The Sub-Millimetre User Reduction Facility; the online manual is located at starlink.eao.hawaii.edu/docs/sun258.htx/sun258.html.

[3] The Spectral Analysis Tool; the online manual is located at astro.dur.ac.uk/ pdraper/splat/sun243.htx/sun243.html.

only, as they are set to be constant over the atmospheric column where absorption arises. Verification tests (against Lincowski et al. 2021, Figure 2 and Greaves et al. 2020, Figure 4) found that our simple code generates $SO_2$ lines that are 15-25% less deep than from more detailed models (see discussion in Appendix A). If our $SO_2$ model-lines are $\approx 0.8\times$ true line-depths, then representative abundances we infer by fitting the depths of observed spectra will be too high by a factor $\approx 1.25$ (i.e. by 1/0.8). This systematic uncertainty is negligible compared to the altitude-dependent variations of molecular abundances in Venus' atmosphere (e.g. Rimmer et al. 2021, and discussion below).

Here we generate our model $SO_2$ spectra at spectral resolutions used in the ALMA and JCMT observations. We adopt as pressure-broadening coefficient the mean-estimate of $0.18$ cm$^{-1}$/atm from Lincowski et al. (2021). The $SO_2$ line-strengths are from Underwood et al. (2016, via the HITRAN database), but our code now additionally[4] scales line-strengths for temperature as a function of altitude, as described in Appendix A. This is important for modelling the "contaminating" $SO_2$ line, as its absorbing energy level is significantly different to that of other transitions that were observed, at 613 K versus 93-152 K. The temperature-pressure profile for Venus' atmosphere is well-defined, with for example a temperature-related uncertainty in the altitude where the 1.1. mm continuum is most strongly generated of $\sim$2-3 km (best-estimate value of 56 km). Since the high-J contaminating $SO_2$ transition will vary the most with adopted temperature of the warm clouds, we tested varying the lowest altitude of absorbing $SO_2$ molecules within these upper/lower bounds. The line-depth of this high-J $SO_2$ transition was found to be uncertain by only $\pm$5% by this method.

For verification of line-ratios, we used the $SO_2$ line-pair of 30(9,21)–31(8,24)/13(3,11)-13(2,12) relevant to the ALMA observations. Detailed codes (Greaves et al. 2020, Lincowski et al. 2021) have given line-depth ratios for this pair of 1:40($\pm$2), while the simplified code gives 1:31. The model used here thus tends to slightly *over*-estimate $SO_2$ (J=31) contamination – i.e. using more detailed calculations would further decrease the negligible contamination we find here.

# 4   RESULTS

## 4.1   ALMA 2019 data

We first note that, after the observatory-led re-processing, absorption near the $PH_3$ 1-0 frequency is clearly detected. Figure 3 shows an example spectrum where we have sought to balance the largest retainable planetary area against the best ripple-cancellation, given that Declination-extremes in the field exhibit worse noise (Greaves et al. 2021a). After further testing of these systematics, we have now extracted a higher signal-to-noise (S/N) detection of $PH_3$ 1-0, from an area that is $\approx$50% larger than in Greaves et al. (2021a – their Supplementary Figure 2). The new extracted spectrum is from a planet-spanning RA,Dec "box", of 97x37 pixels where 1 pixel = 0.16 arcseconds. The long axis of the box is nearly aligned with the planetary equator (which was tilted from RA by 13°), and 49% of the planetary disc is included (no latitudes above 40°).

This spectrum has 7.7 sigma confidence when line-integrated over $\pm$6 km/s. The line-centroid is at -0.2 $\pm$ 0.5 km/s in the Venus-centric $PH_3$ 1-0 reference-frame, fully in agreement with zero as expected if the absorption is dominated by phosphine. This measured velocity is however 3-sigma discrepant with the alternative attribution (Lincowski et al. 2021, e.g.) to $SO_2$ 30(9,21)–3(18,24) absorption, which would be centred in this frame at +1.3 km/s.

We also checked whether the choice of a particular box affected the results. Test-extractions were made for $PH_3$ for bands of latitude (instead of Declination) that extended to 20, 30, 40 or 50 degrees from the equator. The fitted $PH_3$ line-depths varied by only $\sim \pm 10\%$, at the level of noise.

Figure 3 also shows the spectrum from the 97x37-pixel box for the $SO_2$ 13(3,11)-13(2,12) transition that ALMA observed simultaneously. Spectral ripple has been supressed as shown in Figure 2. The overlaid grey curve employs 3 ppb of $SO_2$ at altitudes $\geq$80 km, to demonstrate that mesospheric sulphur dioxide at greater abundances would likely have been detected (at $\geq$3-sigma at line-minimum). We then ran the same model for the $SO_2$ 30(9,21)–3(18,24) transition. Comparing this model to the $PH_3$ 1-0 observation, the $SO_2$ contamination is <2%, when integrated over $\pm$10 km/s (Figure 3b).

## 4.2   JCMT 2017 data

The JCMT acquired $SO_2$ 19(1,19)-18(0,18) spectra on mornings 26, 17 and 6 days before the mid-point of the $PH_3$ run. The $SO_2$ line-continuum ratios were found to be similar, at -0.0043, -0.0029 and -0.0044 at line-minimum respectively, suggesting an $SO_2$ abundance variation of $\sim \pm 20\%$ about the mean. Figure 4 shows the $SO_2$ spectrum observed on 6 June 2017, closest in time to the $PH_3$ observing run. There is some spectral ripple present, for which a suitable least-parameters approximation is a $4^{th}$-order polynomial. After subtracting this trendline from the data, we overlay two line models: one best matching the line core and one with a better match to the wings. The former model is for 20 ppb of $SO_2$ at $\geq$75 km and the latter for 25 ppb at $\geq$80 km. These fits have reduced-$\chi^2$ of 1.8 and 1.65 respectively, over $\pm$20 km/s in velocity – they are representative only (i.e. use constant abundance), but within these limitations the fits match the observed line adequately, and net-abundance and minimum-altitude are quite tightly constrained.

We then generated models for the contaminating transition $SO_2$ 30(9,21)–31(8,24) using these two abundance profiles. Quadratic baseline-fits were subtracted from these models, over the same $\pm$8 km/s region as for the data. (Greaves et al. (2020) applied a higher 8th-order to the data; duplicating this for the model would further reduce the inferred contamination.) Figure 4 shows the results, demonstrating that the $SO_2$(J=31) line-contribution must be small. The model signal, over $\pm$8 km/s, is only 9-11% of the total absorption seen in the $PH_3$ 1-0 observation. The $PH_3$ 1-0 detection shown has 6.7-sigma confidence when integrated over $\pm$8 km/s.

Given the rather constant $SO_2$ that JCMT observed in May-Jun 2017, it seems unlikely that the planet-wide abundance soared by an order of magnitude in the three days between the last $SO_2$ spectrum and the start of $PH_3$ observing. Large jumps in $SO_2$ are rare for big areas of Venus (Marcq et al. 2020). An extreme example is a 2.5-fold decline seen by HST over 5 days (Jessup et al. 2015), but this was restricted to latitudes inside $\pm 10°$, whereas the JCMT $SO_2$ spectrum shown here represents the whole planet and a ten-fold increase would be needed. We conclude that while $SO_2$ does vary strongly on various cadences, the $SO_2$ and $PH_3$ JCMT spectra here are sufficiently contemporaneous for a robust contamination test.

---

[4] This temperature dependence was not included in Greaves et al. (2021a).

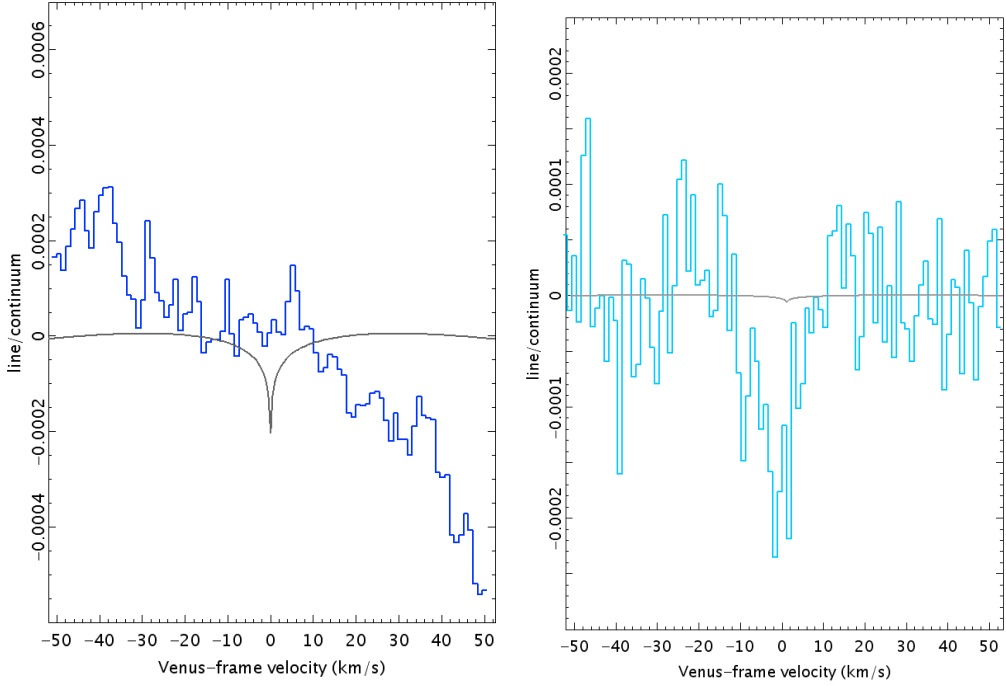

**Figure 3.** Left panel, (a): ALMA wideband spectrum of SO$_2$ 13(3,11)-13(2,12) (dark blue histogram), extracted from half of the planetary disc as described in the text. Spectral ripple has been suppressed as shown in Figure 2. The grey curve is an upper-limit model, for 3 ppb of SO$_2$ at all altitudes ≥80 km, after removal of a quadratic baseline fitted outside ±10 km/s, to simulate the processing of the real data. Right panel, (b): spectrum at the PH$_3$ 1-0 frequency (blue histogram) for the same planetary area. The upper grey histogram is the corresponding (3 ppb, ≥ 80 km, quadratic-subtracted) model demonstrating the upper-limit on the SO$_2$ 30(9,21)–3(18,24) contamination. The grey histogram encloses <2% of the line-area of the blue histogram.

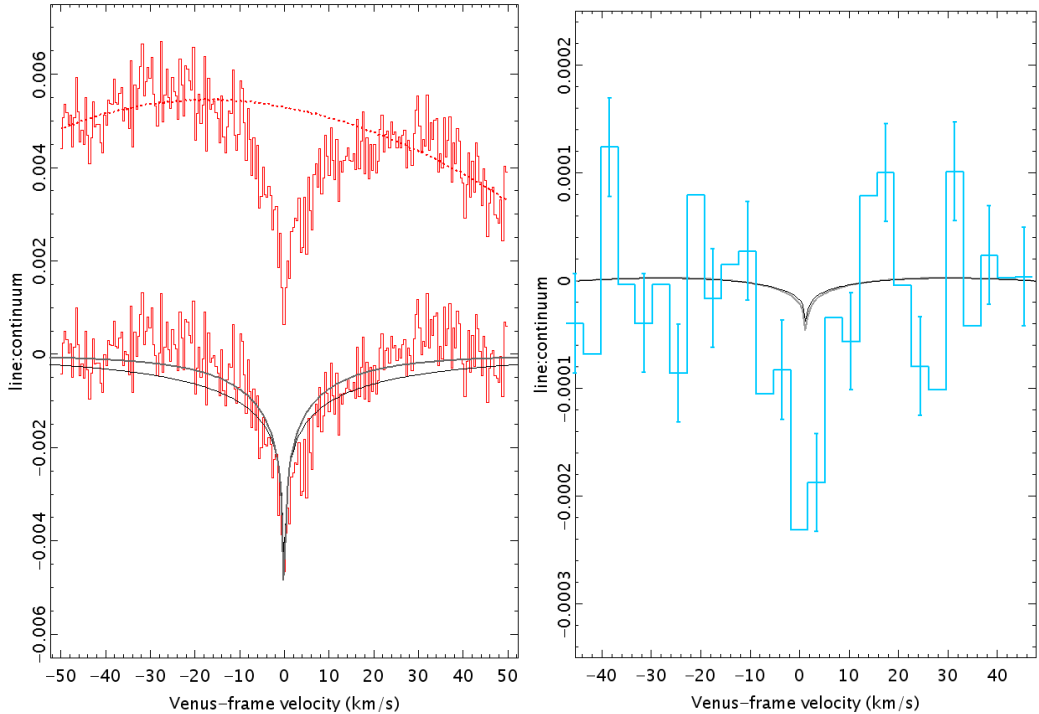

**Figure 4.** Left panel, (a): JCMT 346.6 GHz spectrum of SO$_2$ 19(1,19)-18(0,18) observed on June 6 2017 (upper red histogram, offset vertically for clarity). The dashed trendline is a 4th-order polynomial fitted outside the line region. The lower histogram shows the trendline-subtracted residual spectrum, overlaid with two models: the thicker curve is for 25 ppb of SO$_2$ at ≥80 km and the thinner (lower) curve is for 20 ppb at ≥75 km. Right panel, (b): the net 9-16 June 2017 JCMT spectrum of PH$_3$ 1-0 (blue histogram, spectrally binned and with representative 1-sigma errorbars), overlaid with the same two models as in (a) but for the contaminant SO$_2$ 30(9,21)–3(18,24) line. Observed and model spectra have here both had line-wings suppressed outside ≥8 km/s (see text).

# 5 DISCUSSION

## 5.1 Contamination

Figures 3 and 4 show that the absorptions proposed as PH$_3$ 1-0 can not realistically be re-attributed to sulphur dioxide. Contamination of the observed spectra by the high-J SO$_2$ transition is inferred to be $\leq$10%, at both epochs. The ALMA-detection presented here has 7.7 sigma confidence, minimally affected by up to 2% SO$_2$ contamination. The JCMT spectrum has 6.7 sigma confidence, so is at $\approx$6.0 sigma after removing 10% contamination.

Lincowski et al. (2021) proposed that high abundances of SO$_2$ could escape detection when larger spatial scales are filtered out in ALMA processing, and Akins et al. (2021) simulated this for a range of 2D gas-distributions. We agree with their conclusions about spatial filtering, but emphasise that this can not enhance SO$_2$ 30(9,21)–31(8,24) relative to SO$_2$ 13(3,11)-13(2,12) absorption. Both line-signals will be similarly filtered, as they arise from the same spatial distribution of gas and the same column of absorbing atmosphere (line-frequencies differ <1%). Therefore, a strong upper limit on the observed SO$_2$ J=13 line imposes a stringent limit on the "contaminant" J=31 line – such that ALMA absorption detected at the PH$_3$ frequency can not be solely from SO$_2$.

The only pathological case that we can identify would involve weighting the SO$_2$ abundance-profile to the warmest gas-layer (where the J=31 level-population is higher), and making this layer particularly patchy (so J=31 spatial-filtering is reduced). Imagining a simple two-layer atmosphere, J=13 absorption could trace both a smooth cold mesosphere (temperatures down to ~175 K) and a warm patchy cloud deck (up to ~300 K). J=31 absorption (from an energy level above 600 K) would be more weighted by the warm cloud-decks, and less of this signal would be filtered out. The result would be a stronger "contaminant" SO$_2$ 30(9,21)–3(18,24) line than in standard models. However, this concept severely conflicts with observations. While millimetre-wavelength absorption-spectra have some sensitivity to the clouds (Piccialli et al. 2017, Figure 7, e.g.), these observations mainly find SO$_2$ in the mesosphere (above ~78-85 km: Lincowski et al. 2021). Further, while this mesospheric SO$_2$ *may* be patchy (Encrenaz et al. 2015), lower altitudes are dominated by large-scale Hadley-cell circulation (Marcq et al. 2020), with SO$_2$ "plumes" rapidly smeared out by super-rotation (Encrenaz et al. 2019). A smoother SO$_2$ distribution at lower altitude is the opposite of what is needed for this enhanced-J=31 scenario to work.

## 5.2 Abundances

The net SO$_2$ abundances inferred here are typical for the gas-columns traced by millimetre-wavelength observations, e.g. the 4-year JCMT-monitoring programme of Sandor et al. (2010). Our 20-25 ppb of SO$_2$ from JCMT in June 2017 is near the mean, $\approx$23 ppb, in the Sandor et al. tabulation. Our $\leq$3 ppb ALMA SO$_2$ upper-limit from March 2019 lies in the lowest decile of their abundance table, but with correction for spatial-filtering the limit could be higher. Akins et al. (2021) show that filtering could reduce the SO$_2$ line-depth by a factor as much as ~30 (in their model D, covering an area similar to that outlined in Figure 1).

While our net-SO$_2$ abundances are normal for millimetre monitoring, abundance-profiles that have been inferred from comprehensive (multi-wavelength and in-situ) datasets tend to increase upwards. Such top-weighting enabled Lincowski et al. (2021) to approximately match our JCMT PH$_3$-frequency spectrum using SO$_2$ alone. However, we can now show that their model – Case D, SO$_2$ rising to ~400 ppb at 100 km altitude – is inconsistent with the newly-retrieved JCMT data. We adapted our simple radiative-transfer code for the SO$_2$ 19(1,19)-18(0,18) transition to use the Case D abundance-profile from Lincowski et al. (2021; we made an empirical fit to the curve in their Figure 1b). The minimum height where absorption against the continuum can occur at this frequency was taken to be 60 km (Piccialli et al. 2017). Figure 5 demonstrates that this ad-hoc model – based on various historical SO$_2$ data – will produce ~10× deeper absorption than was actually seen at JCMT. Hence the ad-hoc model greatly over-estimates the amounts of SO$_2$ that were present near the time of the JCMT observations of PH$_3$.

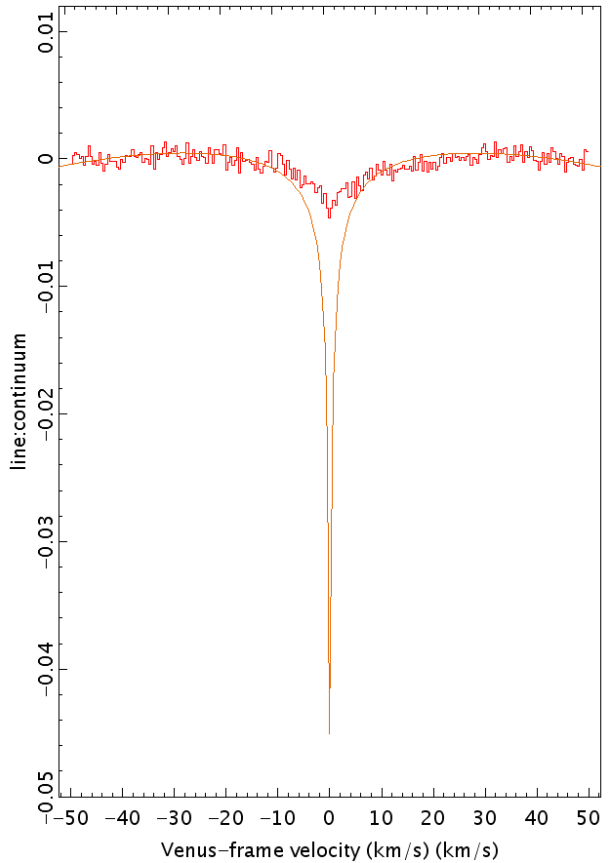

**Figure 5.** Test of a published ad-hoc model for SO$_2$. The red histogram is SO$_2$ 19(1,19)-18(0,18) as observed at JCMT, from Figure 4a. The red curve is a prediction from a SO$_2$ model based on Lincowski et al. (2021), as described in the text. A quadratic was subtracted from the model to remove line wings, similar to the data-processing of Figure 4a.

Even after removing SO$_2$ contamination, it is still problematic to infer PH$_3$ abundances (assuming that the millimetre-absorber is phosphine, and not some mysterious uncatalogued molecule). Villanueva et al. (2021) argue that the narrow observed lines suggest PH$_3$ must be solely mesospheric, and so of high abundance given a limited gas-column to populate. We agree that our preliminary constant-abundance profile (Greaves et al. 2020) was not in fact suited to matching the observed spectra (discussed in Greaves et al. 2021a). Also, a purely mesospheric location is indeed problematic, because PH$_3$ molecules should rapidly photodissociate (unless upwards diffusion is much faster than expected: Greaves et al. 2021a). However, the cloud-level PH$_3$ contribution is unquantified here, given that line-wings are heavily suppressed in necessary processing. The Pioneer-Venus in-cloud (Mogul et al. 2021) support at ~5-sigma confidence that PH₃ is indeed present in the clouds, likely in the mid-to-high ppb range (Bains et al. 2021c). More work is needed to reconcile all available PH₃ data with a consistent profile of abundance by altitude.

We highlight that, although abundances remain unknown due to poor altitude information, the line:continuum ratios for JCMT and ALMA are now similar. After subtracting contamination and adopting uniform ≈3.5 km/s spectral binning, the line-minimum for JCMT is -2.1e-4 and for ALMA it is -1.7e-4, i.e. a variation of ~10% around the mean. An important factor in this convergence is the assumed continuum in calculating the line/continuum depths. In the ALMA case, we are now using the observed continuum signal (which was always the case for JCMT). The true ALMA line-depth is uncertain due to the unknown effects of spatial filtering; in particular, the absorption could be significantly less deep with a Venus-model continuum rather than the observed filtered continuum. This effect was modelled by Akins et al. (2021), while Greaves et al. (2021a) calculated from the data that 22% of the continuum survives the spatial filtering. Modulo this uncertainty, ALMA and JCMT both sampled large areas of Venus (Table 1), and so the similar spectra hint that phosphine-attributed abundances were broadly comparable in 2017 and 2019.

## 6 CONCLUSIONS

(Quasi)-simultaneous data on sulphur dioxide absorption are used to infer bounds on the SO₂ contamination of absorption at the PH₃ 1-0 frequency. These limits are <2% to ≈10%, for large areas of Venus. Sulphur dioxide would need to have increased ten-fold planet-wide over only a few days for SO₂ to have mimicked PH₃ in the discovery data. We conclude that results from observations preclude re-attributing phosphine on Venus to sulphur dioxide.

## ACKNOWLEDGEMENTS

We thank an anonymous referee for comments that greatly helped in improving the clarity of the manuscript. JCMT data presented here were obtained under program id M17AP081 (retrieved from the archive hosted at CADC) and for the authors' service-program S16BP007. We thank the PI of M17AP081, Hideo Sagawa, for valuable consultations about interpretation of these observations. The James Clerk Maxwell Telescope is operated by the East Asian Observatory on behalf of The National Astronomical Observatory of Japan; Academia Sinica Institute of Astronomy and Astrophysics; the Korea Astronomy and Space Science Institute; Center for Astronomical Mega-Science (as well as the National Key R&D Program of China with No. 2017YFA0402700). Additional funding support is provided by the Science and Technology Facilities Council of the United Kingdom and participating universities and organizations in the United Kingdom and Canada. This paper makes use of the following ALMA data: ADS/JAO.ALMA#2018.A.00023.S. ALMA is a partnership of ESO (representing its member states), NSF (USA) and NINS (Japan), together with NRC (Canada), MOST and ASIAA (Taiwan), and KASI (Republic of Korea), in cooperation with the Republic of Chile. The Joint ALMA Observatory is operated by ESO, AUI/NRAO and NAOJ. We thank staff at ESO for their speedy and substantial efforts in re-processing the ALMA data. This research used the facilities of the Canadian Astronomy Data Centre operated by the National Research Council of Canada with the support of the Canadian Space Agency. SMURF and SPLAT software were provided by the UK Starlink Project, via http://www.starlink.ac.uk/docs/starlinksummary.html.

## APPENDIX A: LINE PROFILE MODEL

We develop a semi-analytical one-dimensional model for the line profile, starting at the surface of Venus, height $z = 0$ km, and moving up in height increments based on the same temperature profile used by Greaves et al. (2021a), with the top of the atmosphere at 115 km.

We track the intensity as a function of height ($z$, km) and frequency ($\nu$, Hz), with a line center at $\nu_0$ (Hz):

$$I = I(\nu, \nu_0, z), \tag{A1}$$

We set $I(\nu, \nu_0, 0)$ equal to unity over all frequencies. For each step, $\Delta z$ (km), from 0 km to 115 km, we update:

$$I(\nu, \nu_0, z + \Delta z) = I(\nu, \nu_0, z)e^{-\tau(z, z+\Delta z)} + \frac{B_\nu(z)}{B_\nu(0)}e^{-\tau(z, z+\Delta z)}, \tag{A2}$$

where $\tau$ is the optical depth between z and $z + \Delta z$, given by:

$$\tau(z, z + \Delta z) = f_V(\nu, \nu_0, z)S^*_{\nu_0}(z)\Delta z. \tag{A3}$$

Here $f_V(\nu, \nu_0, z)$ is the Voigt profile, and depends on the gas temperature $T$ (K) and pressure $p$ (bar), both functions of height, $z$, and on the pressure-broadening coefficient $\gamma$ (cm$^{-1}$/atm). The line-strength, $S^*_{\nu_0}$ (cm$^{-1}$ s$^{-1}$) is:

$$S^*_{\nu_0}(z) = cn(z)e^{E_J/T_0}e^{-E_J/T(z)}S_{\nu_0}, \tag{A4}$$

where c = $2.9979 \times 10^{10}$ cm/s is the speed of light, $n(z)$ (cm$^{-3}$) is the number density of the absorbing species, $E_J$ (K) is the value of the lower energy level, $T$ (K) is the temperature and $S_{\nu_0}$ (cm$^{-1}$/(molecules cm$^{-2}$) is the line strength at $T_0 = 300$ K. The reason for the two Boltzmann terms is to scale $S_{\nu_0}$ to different temperatures. We assume that the atmospheric gas at each height is in local thermodynamic equilibrium and therefore the population of the lower level is proportional to $e^{-E_J/T}$. We also assume that the gas in the experiment used to measure $S_{\nu_0}$ at $T_0$ is also in local thermodynamic equilibrium, and therefore is proportional to $e^{-E_J/T_0}$. We can achieve the line strength appropriate for $T$ by dividing the former Boltzmann factor by the latter. Finally $B_\nu(z)$ is the spectral radiance given by Planck's law:

$$B_\nu(z) = \frac{2h\nu^3}{c^2}\frac{1}{e^{h\nu/(kT)} - 1}, \tag{A5}$$

where h = $6.62607 \times 10^{-27}$ erg s is Planck's constant and k = $1.38065 \times 10^{-16}$ erg/K is Boltzmann's constant.

This one-dimensional model accounts for absorption and thermal emission, but does not treat scattering. Scattering will increase the path length traversed and is therefore expected to strengthen the absorption features. Unless scattering is steeply dependent on frequency, this effect will be approximately the same for each line, and so the ratio of line strengths will not be affected.

This program calculates a single line profile and is written in Python3. Example code to model the SO$_2$ 13(3,11)-13(2,12) transition can be found at http://www.mrao.cam.ac.uk/~pbr27/J13.py.

This paper has been typeset from a TeX/LaTeX file prepared by the author.